\documentclass[12pt]{iopart}

\usepackage{amsfonts,cancel}

\newcommand{\Lie}{\mathcal{L}}
\newcommand{\cg}{\bar \gamma}
\newcommand{\cK}{\bar K}

\begin{document}

\title{Potentials for transverse trace-free tensors}

\author{Rory Conboye and Niall \'{O} Murchadha}

\address{Physics Department, University College, Cork, Ireland}
\ead{rconboye@gmail.com; niall@ucc.ie}

\begin{abstract}
In the initial conditions of the $3 + 1$ formalism for numerical relativity, the transverse and trace-free (TT) part of the extrinsic curvature plays a key role. We know that TT tensors possess two degrees of freedom per space point.   However, finding an expression for a TT tensor depending on only two scalar functions is a non-trivial task. Assuming either axial or translational  symmetry, expressions depending on two scalar potentials alone are derived here for \emph{all} TT tensors in flat $3$-space. In a more general spatial slice, only one of these potentials is found, the same potential given in \cite{BakerPuzio} and \cite{Dain}, with the remaining equations reduced to a partial differential equation, depending on boundary conditions for a solution. As an exercise, we also derive the potentials which give the Bowen-York curvature tensor in flat space.
\end{abstract}

\section{Introduction}

This article is devoted to TT tensors. These are symmetric $2$-index tensors which are both transverse (divergence-free) and trace free. Therefore one has $4$ conditions on $6$ variables, and expects TT tensors to have $2$ degrees of freedom per space point. In this article we consider  TT tensors on flat space and assume that the tensor has an extra symmetry, either translational or rotational. Under these assumptions we can express all such TT tensors in terms of two free scalar potentials.

In the $3 + 1$ formalism for numerical relativity, the initial conditions are given by a space-like hypersurface, with a $3$-space metric $\gamma_{a b}$ and an extrinsic curvature tensor $K^{a b}$, giving the embedding of the hypersurface in the space-time manifold. The decomposition of the Einstein equations however, gives a set of constraint equations for the metric and curvature, restricting the choices for $4$ of the $12$ components (both $\gamma_{a b}$ and $K^{a b}$ being symmetric $3$-space tensors). In the weak-field regime, \cite{ADM}, it was realised early that the true degrees of freedom consisted of a pair of TT tensors on flat space. Thus we have the desired $4$ degrees of freedom, once the $4$ degreees associated with the choice of coordinates are removed. One can see an immediate parallel with Maxwell's equations, where the degrees of freedom consists of two divergence-free vectors.

Even in the strong-field regime, the most successful technique for dealing with which components to constrain, and which to treat as free initial data, has come from the conformal transverse trace-free (CTT) decomposition, originally developed in \cite{York71}, \cite{OMYork}, \cite{YorkTT} (see also the text books \cite{Alc} and \cite{BS}). In the CTT decomposition, a conformal transformation of the TT part of the extrinsic curvature, with respect to a conformally transformed spatial metric, is generally taken to represent two of the component freedoms of the initial conditions.

Unfortunately, explicit expressions for a TT tensor, depending on two scalar functions alone, cannot be given in general, with the form of the tensor also dependent on boundary conditions. However, with the existence of a surface-orthogonal Killing field, a single scalar potential is found in \cite{BakerPuzio}, with further input required for the rest of the tensor. This result is formulated in a coordinate-independent manner in \cite{Dain}, using a ``time/Killing-coordinate symmetry'' from \cite{BrandtSeidel}, which sends the remaining components to zero.

Here these results are developed further, with explicit expressions found for TT tensors in flat space, depending entirely on two scalar potentials. These are derived in \sref{sec:TTT Flat Space Linear} in both Cartesian and cylindrical coordinates, with a linear symmetry condition, and in \sref{sec:TTT Flat Space Axial} in cylindrical and spherical coordinates, with the condition of axial symmetry. For a more general space, the scalar potential found in both \cite{BakerPuzio} and \cite{Dain} is derived in \sref{sec:TTT General Space} in cyindrical coordinates, for axially-symmetric tensors. Though a second potential cannot be found explicitly, the remaining equations are reduced to a second order partial differential equation in two of the remaining components. For the flat axially-symmetric tensors already derived, the choice of potentials which give the Bowen-York curvature tensor \cite{BY} are computed in section \ref{sec:TTT BY}.

Throughout the paper, the existence of the spatial integrals of the tensor components is assumed, and in most cases double spatial integrals.

\section{Flat Space TT Tensors with Linear Symmetry}
\label{sec:TTT Flat Space Linear}

 We  begin with linear symmetry.  Of course, this is a much less interesting case than axial symmetry, because we lose the possibility of asymptotic flatness. We do this because the calculations are easier and also help us understand what needs be done in the axial symmetry case. The calculations are much easier if one uses coordinates which reflect the underlying symmetry, and with translational symmetry, the natural coordinates are Cartesian and cylindrical polars. Transverse and trace-free tensor expressions are thus derived in both Cartesian and cylindrical coordinates. As desired, the Cartesian expression is dependent on two scalar potentials alone, while the cylindrical expression contains an integral, as well as a function of integration. A coordinate transformation is performed to relate the potentials in the two coordinate systems, showing how to write an expression in cylindrical coordinates which has no integral and no function of integration.

\subsection{Cartesian Coordinates}
\label{sec:TT F Cartesian}

In Cartesian coordinates $(x, y, z)$, the flat $3$-space line element is given simply by:
\begin{equation}\label{TT FCar Metric}
d l^2 \ = \ d x^2 + d y^2 + d z^2 \ ,
\end{equation}
with completely vanishing connection coefficients. Now, a symmetric two-index tensor $T^{a b}$ is invariant along a Killing vector field $\eta^a$, if and only if its Lie derivative with respect to that vector field is zero:
\begin{equation}
0 \
  = \ \Lie_\eta T^{a b}
  = \ \eta^c \partial_c T^{a b}
    - T^{c b} \partial_c \eta^a
    - T^{a c} \partial_c \eta^b
        \ .
        \label{TT F Lie Tab}
\end{equation}
Taking the Killing vector to coincide with the coordinate vector $z$, \eref{TT F Lie Tab} reduces to:
\begin{equation}\label{TT FL partial z Tab}
0 \
 =  \ z^c \partial_c T^{a b}
    - T^{c b} \cancel{\partial_c z^a}
    - T^{a c} \cancel{\partial_c z^b} \
 \equiv \ \partial_z T^{a b} \ ,
\end{equation}
giving a simple condition for $T^{a b}$ to be linearly-symmetric. Combining this condition with the equations for $T^{a b}$ to be transverse and trace-free:
\numparts
\begin{eqnarray}
0 \
 =  \ D_b T^{x b}
 =    \partial_x T^{x x}
    + \partial_y T^{x y}
    + \cancel{\partial_z T^{x z}} \ ,
        \label{TT FCar DivTx0} \\
0 \
 =  \ D_b T^{y b}
 =    \partial_x T^{y x}
    + \partial_y T^{y y}
    + \cancel{\partial_z T^{y z}} ,
        \label{TT FCar DivTy0} \\
0 \
 =  \ D_b T^{z b}
 =    \partial_x T^{z x}
    + \partial_y T^{z y}
    + \cancel{\partial_z T^{z z}} ,
	\label{TT FCar DivTz0} \\
0 \
 =  \ T^{x x}
    + T^{y y}
    + T^{z z} .
      \label{TT FCar trT0}
\end{eqnarray}
\endnumparts
By the equivalence of mixed partial derivatives, \eref{TT FCar DivTx0}, \eref{TT FCar DivTy0} and \eref{TT FCar DivTz0} then imply the existence of scalar potentials $P$, $Q$ and $S$, with:
\begin{eqnarray}
- \partial_x T^{x x} \
 =  \ \partial_y T^{x y}
	\qquad &\Leftrightarrow \qquad
T^{x x} \
 =  \ - \partial_y P \ ,
	\qquad
T^{x y} \
 =  \ \partial_x P
	\ , \label{TT FCar P} \\
- \partial_x T^{y x}
 =  \ \partial_y T^{y y}
	\qquad &\Leftrightarrow \qquad
T^{y x} \
 =  \ - \partial_y Q \ , \qquad
T^{y y} \
 =  \ \partial_x Q
	\ , \label{TT FCar Q} \\
- \partial_x T^{z x}
 =  \ \partial_y T^{z y}
	\qquad &\Leftrightarrow \qquad
T^{x z} \
 =  \ - \partial_y S \ , \qquad
T^{y z} \
 =  \ \partial_x S
	\ . \label{TT FCar S}
\end{eqnarray}
Also, since $T^{a b}$ is symmetric, the expressions for $T^{x y}$ in \eref{TT FCar P} and $T^{y x}$ in \eref{TT FCar Q} can be equated, implying the existence of another potential $R$, with:
\begin{equation}
\partial_x P \
 =  \ - \partial_y Q
	\qquad \Leftrightarrow \qquad
P \
 =  \ \partial_y R \ , \qquad
Q \
 =  \ - \partial_x R \ . \label{TT FCar R}
\end{equation}
Substituting \eref{TT FCar R} into \eref{TT FCar P} and \eref{TT FCar Q}, and using \eref{TT FCar trT0} to find $T^{z z}$, the tensor $T^{a b}$ is given in matrix form by:
\begin{equation}\label{TT FCar}
T^{a b} \ = \
\left(
  \begin{array}{ccc}
     - \partial_{y y} R
     & \partial_{x y} R
     & - \partial_y S \\
   & & \\
       \partial_{x y} R
     & - \partial_{x x} R
     & \partial_x S \\
   & & \\
       - \partial_y S
     & \partial_x S
     & \partial_{x x} R + \partial_{y y} R \\
  \end{array}
 \right) \ ,
\end{equation}
depending on the choice of the two scalar potentials $R$ and $S$ alone. Of course, it is assumed that $R$ and $S$ are independent of $z$. This expression is then transverse, trace-free, and linearly-symmetric along the $z$ coordinate.

\subsection{Cylindrical Coordinates}
\label{sec:TT FLC}

In cylindrical coordinates $(\rho, \phi, z)$, the flat $3$-space line element is given by:
\begin{equation}\label{TT FLC Metric}
d l^2 \ = \ d \rho^2 + \rho^2 d \phi^2 + d z^2 \ ,
\end{equation}
with non-zero connection coefficients:
\begin{equation}\label{TT FLC Connections}
     \Gamma^\rho_{\phi \phi} \
 = \ - \rho \ , \qquad
     \Gamma^\phi_{\rho \phi} \
 = \ \Gamma^\phi_{\phi \rho} \
 = \ \frac{1}{\rho} \ .
\end{equation}
With the divergence of a tensor $T^{a b}$ given by:
\begin{equation}\label{Divergence of Tab}
D_b \ T^{a b} \
 =  \ \partial_b \ T^{a b}
    + \Gamma^{a}_{b c} \ T^{c b}
    + \Gamma^{b}_{b c} \ T^{a c} \ ,
\end{equation}
the transverse and trace-free conditions for a symmetric tensor $T^{a b}$, along with the condition of linear symmetry along the $z$ coordinate from \eref{TT FL partial z Tab}, are given by the equations:
\numparts
\begin{eqnarray}
0
 =  D_b T^{\rho b}
 =    \partial_\rho T^{\rho \rho}
    + \cancel{\partial_z T^{\rho z}}
    + \partial_\phi T^{\rho \phi}
    - \rho \ T^{\phi \phi}
    + \frac{1}{\rho} \ T^{\rho \rho} ,
        \label{TT FLC DivTrho0} \\
0
 =  D_b T^{\phi b}
 =    \partial_\rho T^{\phi \rho}
    + \cancel{\partial_z T^{\phi z}}
    + \partial_\phi T^{\phi \phi}
    + \frac{3}{\rho} \ T^{\rho \phi} ,
	\label{TT FLC DivTphi0} \\
0
 =  D_b T^{z b}
 =    \partial_\rho T^{z \rho}
    + \cancel{\partial_z T^{z z}}
    + \partial_\phi T^{z \phi}
    + \frac{1}{\rho} \ T^{z \rho} ,
        \label{TT FLC DivTz0} \\
0
 =  T^{\rho \rho}
    + T^{z z}
    + \rho^2 \ T^{\phi \phi} .
      \label{TT FLC trT0}
\end{eqnarray}
\endnumparts

Firstly, \eref{TT FLC DivTz0} is manipulated:
\begin{eqnarray}
&0
 &= \frac{1}{\rho} \
      \partial_\rho \left(\rho \ T^{z \rho}\right)
    + \partial_\phi \left(T^{z \phi}\right) \nonumber \\
 &&= \frac{1}{\rho} \
      \partial_\rho \left(\rho \ T^{z \rho}\right)
    + \frac{1}{\rho} \
      \partial_\phi \left(\rho \ T^{z \phi}\right) \ , \nonumber \\
	\nonumber \\
\Leftrightarrow \qquad
    &&\partial_\rho \left(- \rho \ T^{z \rho}\right) \
 = \ \partial_\phi \left(\rho \ T^{z \phi}\right) \ ,
	\label{TT FLC Tzr-Tzz}
\end{eqnarray}
and by the equivalence of mixed partial derivatives, there must exist a scalar potential $Y_L$, such that:
\begin{eqnarray}
&\partial_\phi Y_L \
 =  \ - \rho \ T^{\rho z} \ , \qquad
&\partial_\rho Y_L \
 =  \ \rho \ T^{z \phi} \ , \nonumber \\
	\nonumber \\
\Leftrightarrow \qquad
&T^{\rho z} \
 =  \ - \frac{1}{\rho} \partial_\phi Y_L \ , \qquad
&T^{z \phi} \
 =  \ \frac{1}{\rho} \partial_\rho Y_L \ .
	\label{TT FLC YL}
\end{eqnarray}

Similarly with \eref{TT FLC DivTphi0}:
\begin{eqnarray}
&0
 &= \frac{1}{\rho^3} \ \partial_\rho
      \left( \rho^3 \ T^{\rho \phi} \right)
    + \partial_\phi \left(T^{\phi \phi}\right) \nonumber \\
 &&= \frac{1}{\rho^3} \ \partial_\rho
      \left( \rho^3 \ T^{\rho \phi} \right)
    + \frac{1}{\rho^3} \ \partial_\phi
      \left( \rho^3 \ T^{\phi \phi} \right) \ , \nonumber \\
	\nonumber \\
\Leftrightarrow \qquad
    &&\partial_\rho
     \left( \rho^3 \ T^{\rho \phi} \right) \
 = \ \partial_\phi
     \left( - \rho^3 \ T^{\phi \phi} \right) \ ,
	\label{TT FLC Trp-Tzp}
\end{eqnarray}
and again, by the equivalence of mixed partial derivatives there must exist a scalar potential $X_L$, such that:
\begin{eqnarray}
&\partial_\phi X_L \
 =  \ \rho^3 \ T^{\rho \phi} \ , \qquad
&\partial_\rho X_L \
 =  \ - \rho^3 \ T^{\phi \phi} \ , \nonumber \\
	\nonumber \\
\Leftrightarrow \qquad
&T^{\rho \phi} \
 =  \ \frac{1}{\rho^3} \ \partial_\phi X_L \ , \qquad
&T^{\phi \phi} \
 =  \ - \frac{1}{\rho^3} \ \partial_\rho X_L \ .
\label{TT FLC XL}
\end{eqnarray}

Finally, beginning with \eref{TT FLC DivTrho0} and substituting from \eref{TT FLC XL}:
\begin{eqnarray}
&0
 &= \partial_\rho T^{\rho \rho}
    + \partial_\phi T^{\rho \phi}
    - \rho \ T^{\phi \phi}
    + \frac{1}{\rho} \ T^{\rho \rho}
	\nonumber \\
 &&= \frac{1}{\rho} \ \partial_\rho
      \left(\rho \ T^{\rho \rho}\right)
    + \frac{1}{\rho^3} \ \partial_{\phi \phi} X_L
    + \frac{1}{\rho^2} \ \partial_\rho X_L
    \ , \nonumber \\
	\nonumber \\
\Leftrightarrow \qquad
&&\partial_\rho \left(\rho \ T^{\rho \rho}\right) \
 =  - \frac{1}{\rho} \ \partial_\rho X_L
    - \frac{1}{\rho^2} \ \partial_{\phi \phi} X_L
    \ , \nonumber \\
	\nonumber \\
\Leftrightarrow \qquad
&&T^{\rho \rho}
 =  - \frac{1}{\rho} \int \left[
        \frac{1}{\rho} \ \partial_\rho X_L
      + \frac{1}{\rho^2} \ \partial_{\phi \phi} X_L
      \right] d \rho
    - \frac{1}{\rho} \ f_L(\phi)
	\ , \label{TT FLC Trr}
\end{eqnarray}
noting the addition of a function of integration, depending on $\phi$, but not $\rho$ (or $z$).

Transverse, trace-free and linearly-symmetric tensors in flat space can then be given, in cylindrical coordinates, by the matrix expression:
\begin{eqnarray}
\fl &T^{a b} \ = \nonumber \\
\fl &\left(
  \begin{array}{ccc}
    - \frac{1}{\rho} \int \left[
        \frac{1}{\rho} \ \partial_\rho X_L
      + \frac{1}{\rho^2} \ \partial_{\phi \phi} X_L
      \right] d \rho
     & \frac{1}{\rho^3} \ \partial_\phi X_L
     & - \frac{1}{\rho} \partial_\phi Y_L \\
    - \frac{1}{\rho} \ f_L(\phi)
   & & \\
   & & \\
    \frac{1}{\rho^3} \ \partial_\phi X_L
     & - \frac{1}{\rho^3} \ \partial_\rho X_L
     & \frac{1}{\rho} \partial_\rho Y_L \\
   & & \\
    - \frac{1}{\rho} \partial_\phi Y_L
     & \frac{1}{\rho} \partial_\rho Y_L
     & \frac{1}{\rho} \ \partial_\rho X_L
       + \frac{1}{\rho} \ f_L(\phi) \\
   & & + \frac{1}{\rho} \int \left[
         \frac{1}{\rho} \ \partial_\rho X_L
       + \frac{1}{\rho^2} \ \partial_{\phi \phi} X_L
       \right] d \rho \\
  \end{array}
 \right) \ , \label{TT FLC}
\end{eqnarray}
with two scalar potentials $X_L$ and $Y_L$, and a function of integration $f_L(\phi)$.

\subsection{Coordinate Transformations}

Since both \eref{TT FCar} and \eref{TT FLC} give expressions for the same type of tensors, but in different coordinate systems, the potentials from each should be related. A coordinate transformation is therefore performed, and since the Cartesian expression depends on the desired two potentials alone, we transform \eref{TT FCar} into cylindrical coordinates.

To begin, the derivatives of $R$ and $S$ with respect $x$ and $y$ are first found in terms of cylindrical coordinates, by use of the chain rule, giving:
\begin{eqnarray}
\partial_{x x} R \
 &= \ \cos^2 \phi \ \partial_{\rho \rho} R
        + \frac{1}{\rho} \sin^2 \phi \ \partial_\rho R
        - \frac{2}{\rho} \cos \phi \sin \phi \ \partial_{\rho \phi} R
	    \nonumber \\ &
        + \frac{2}{\rho^2} \cos \phi \sin \phi \ \partial_\phi R
        + \frac{1}{\rho^2} \sin^2 \phi \ \partial_{\phi \phi} R
    \ , \nonumber \\
& \nonumber \\
\partial_{x y} R \
 &= \ \cos \phi \sin \phi \ \partial_{\rho \rho} R
        - \frac{1}{\rho} \cos \phi \sin \phi \ \partial_\rho R
        + \frac{1}{\rho} \left(
            \cos^2 \phi - \sin^2 \phi
          \right) \ \partial_{\rho \phi} R
	    \nonumber \\ &
        - \frac{1}{\rho^2} \left(
            \cos^2 \phi - \sin^2 \phi
          \right) \ \partial_\phi R
        - \frac{1}{\rho^2} \cos \phi \sin \phi \ \partial_{\phi \phi} R
    \ , \nonumber \\
& \nonumber \\
\partial_{y y} R \
 &= \ \sin^2 \phi \ \partial_{\rho \rho} R
        + \frac{1}{\rho} \cos^2 \phi \ \partial_\rho R
        + \frac{2}{\rho} \cos \phi \sin \phi \partial_{\rho \phi} R
	    \nonumber \\ &
        - \frac{2}{\rho^2} \cos \phi \sin \phi \partial_\phi R
        + \frac{1}{\rho^2} \cos^2 \phi \ \partial_{\phi \phi} R
    \ , \nonumber \\
& \nonumber \\
\partial_x S \
 &= \ \cos \phi \ \partial_\rho S
        - \frac{1}{\rho} \sin \phi \ \partial_\phi S
    \ , \nonumber \\
\partial_y S 
 &= \ \sin \phi \ \partial_\rho S
        + \frac{1}{\rho} \cos \phi \ \partial_\phi S
    \ .
\end{eqnarray}
These expressions are then substituted into the transformation of \eref{TT FCar} from Cartesian into cylindrical coordinates:
\begin{eqnarray}
\fl T^{\rho \rho} \
 &= \ - \cos^2 \phi \ \partial_{y y} R
    + 2 \cos \phi \sin \phi \ \partial_{x y} R
    - \sin^2 \phi \ \partial_{x x} R
  = \ - \frac{1}{\rho} \ \partial_\rho R
      - \frac{1}{\rho^2} \ \partial_{\phi \phi} R
	\ , \nonumber \\
\fl & \nonumber \\
\fl T^{\phi \phi} \ 
 &= \ - \frac{\sin^2 \phi}{\rho^2} \ \partial_{y y} R
    - \frac{2 \cos \phi \sin \phi}{\rho^2} \ \partial_{x y} R
    - \frac{\cos^2 \phi}{\rho^2} \ \partial_{x x} R
  = \ - \frac{1}{\rho^2} \ \partial_{\rho \rho} R
	\ , \nonumber \\
\fl & \nonumber \\
\fl T^{\rho \phi} \
 &= \ \frac{\cos \phi \sin \phi}{\rho} \ \partial_{y y} R
    + \left(
        \frac{\cos^2 \phi}{\rho}
      - \frac{\sin^2 \phi}{\rho}
      \right) \ \partial_{x y} R
    - \frac{\cos \phi \sin \phi}{\rho} \ \partial_{x x} R
	\nonumber \\ \fl
 &= \ + \frac{1}{\rho^2} \ \partial_{\rho \phi} R
      - \frac{1}{\rho^3} \ \partial_\phi R
	\ , \nonumber \\
\fl & \nonumber \\
\fl T^{z z} \
 &= \ \partial_{x x} R + \partial_{y y} R
  = \   \partial_{\rho \rho} R
      + \frac{1}{\rho} \ \partial_\rho R
      + \frac{1}{\rho^2} \ \partial_{\phi \phi} R
	\ , \nonumber \\
\fl & \nonumber \\
\fl T^{\rho z} \
 &= \ - \cos \phi \ \partial_y S
    + \sin \phi \ \partial_x S \
  = \ - \frac{1}{\rho} \ \partial_\phi S
	\ , \nonumber \\
\fl & \nonumber \\
\fl T^{\phi z} \
 &= \ \frac{\sin \phi}{\rho} \ \partial_y S
    + \frac{\cos \phi}{\rho} \ \partial_x S \
  = \ \frac{1}{\rho} \ \partial_\rho S
	\ ,
\end{eqnarray}
giving a new expression for a linearly symmetric TT tensor in cylindrical coordinates. In matrix form:
\begin{equation}
\fl T^{a b} =
\left(
  \begin{array}{ccc}
    - \frac{1}{\rho} \ \partial_\rho R
      - \frac{1}{\rho^2} \ \partial_{\phi \phi} R
     & \frac{1}{\rho^2} \ \partial_{\rho \phi} R
      - \frac{1}{\rho^3} \ \partial_\phi R
     & - \frac{1}{\rho} \ \partial_\phi S \\
   & & \\
    \frac{1}{\rho^2} \ \partial_{\rho \phi} R
      - \frac{1}{\rho^3} \ \partial_\phi R
     & - \frac{1}{\rho^2} \ \partial_{\rho \rho} R
     & \frac{1}{\rho} \ \partial_\rho S \\
   & & \\
    - \frac{1}{\rho} \ \partial_\phi S
     & \frac{1}{\rho} \ \partial_\rho S
     & \partial_{\rho \rho} R
      + \frac{1}{\rho} \ \partial_\rho R
      + \frac{1}{\rho^2} \ \partial_{\phi \phi} R \\
  \end{array}
 \right) \ , \label{TT FLC R,S}
\end{equation}
dependending only on the two scalar potentials $R$ and $S$, with no funtion of integration.

Comparing \eref{TT FLC} with \eref{TT FLC R,S}, the $T^{\rho z}$ and $T^{\phi z}$ terms can easily be seen to give an equivalence between the potentials $Y_L$ and $S$. Equating each of the remaining terms, and using integration by parts, an expression can also be found for the potential $X_L$ in terms of $R$:
\begin{equation}\label{TT FL X,Y - R,S}
X_L \ = \ \rho \ \partial_\rho R - R \ ,
 \hspace{1cm}
Y_L \ = \ S \ .
\end{equation}
There is also a general solution for the ordinary differential equation in $R$, giving:
\begin{equation}
R \
  = \ \rho \int \frac{1}{\rho^2} \ X_L \ d \rho
    + \rho \ h_L (\phi)
	\ . \label{TT FL R(X)}
\end{equation}
Note also the function of integration $h_L(\phi)$.

\section{Flat Space TT Tensors with Axial Symmetry}
\label{sec:TTT Flat Space Axial}

The techniques of section \ref{sec:TTT Flat Space Linear} are used in this section to derive expressions for an \emph{axially}-symmetric TT tensor, in both cylindrical and spherical coordinates. A similar relation to \eref{TT FL X,Y - R,S} is also found, transforming the tensor in cylindrical coordinates into an integral-free expression. A coordinate transformation is then carried out on this expression, to allow us do the same in spherical coordinates.

\subsection{Cylindrical Coordinates}
\label{sec:TT FAC}

In this section, the cylindrical coordinates are given in the order $(\rho, \ z, \ \phi)$, for easy comparison with the spherical coordinates $(r, \ \theta, \ \phi)$. However it must be noted that this order produces a ``left hand orthogonality'', reversing the sign of the Levi-Civita tensor. The flat $3$-space line element, in these coordinates, is given by:
\begin{equation}\label{TT FAC Metric}
d l^2 \ = \ d \rho^2 + d z^2 + \rho^2 d \phi^2 \ ,
\end{equation}
with its non-zero connection coefficients, as with \eref{TT FLC Connections}:
\begin{equation}\label{TT FAC Connections}
     \Gamma^\rho_{\phi \phi} \
 = \ - \rho \ , \qquad
     \Gamma^\phi_{\rho \phi} \
 = \ \Gamma^\phi_{\phi \rho} \
 = \ \frac{1}{\rho} \ .
\end{equation}

For a symmetric two-index tensor $T^{a b}$ to be axially symmetric, the Killing vector field $\eta^a$ from \eref{TT F Lie Tab} can be taken to coincide with the azimuthal coordinate vector $\phi$, giving the condition:
\begin{equation}\label{TT FA partial phi Tab}
0 \
 =  \ \phi^c \partial_c T^{a b}
    - T^{c b} \cancel{\partial_c \phi^a}
    - T^{a c} \cancel{\partial_c \phi^b} \
 \equiv \ \partial_\phi T^{a b} \ .
\end{equation}
The transverse and trace-free conditions for a symmetric tensor $T^{a b}$, along with the condition of axial symmetry, are then given by the equations:
\numparts
\begin{eqnarray}
0
 =  D_b T^{\rho b}
 =    \partial_\rho T^{\rho \rho}
    + \partial_z T^{\rho z}
    + \cancel{\partial_\phi T^{\rho \phi}}
    - \rho \ T^{\phi \phi}
    + \frac{1}{\rho} \ T^{\rho \rho} ,
        \label{TT FAC DivTrho0} \\
0
 =  D_b T^{z b}
 =    \partial_\rho T^{z \rho}
    + \partial_z T^{z z}
    + \cancel{\partial_\phi T^{z \phi}}
    + \frac{1}{\rho} \ T^{z \rho} ,
        \label{TT FAC DivTz0} \\
0
 =  D_b T^{\phi b}
 =    \partial_\rho T^{\phi \rho}
    + \partial_z T^{\phi z}
    + \cancel{\partial_\phi T^{\phi \phi}}
    + \frac{3}{\rho} \ T^{\rho \phi} ,
	\label{TT FAC DivTphi0} \\
0
 =  T^{\rho \rho}
    + T^{z z}
    + \rho^2 \ T^{\phi \phi} .
      \label{TT FAC trT0}
\end{eqnarray}
\endnumparts
By the equivalence of mixed partial derivatives, \eref{TT FAC DivTz0} and \eref{TT FAC DivTphi0} respectively imply the existence of scalar functions $X_A$ and $Y_A$, with:
\begin{eqnarray}
\fl \partial_\rho \left(\rho \ T^{z \rho}\right) \
 = \ \partial_z \left(- \rho \ T^{z z}\right)
	\ \  &\Leftrightarrow  \qquad
\partial_z X_A \
 =  \ \rho \ T^{\rho z} \ , \qquad
&\partial_\rho X_A \
 =  \ - \rho \ T^{z z} \ ,  \nonumber \\
		&\Leftrightarrow \qquad
T^{\rho z} \
 =  \ \frac{1}{\rho} \partial_z X_A \ , \qquad
&T^{z z} \
 =  \ - \frac{1}{\rho} \partial_\rho X_A \ ,
	\label{TT FAC X} \\
&& \nonumber \\
\fl \partial_\rho
     \left(\rho^3 \ T^{\rho \phi} \right) \
 = \ \partial_z
     \left(- \rho^3 \ T^{z \phi} \right)
	\ \ &\Leftrightarrow \qquad
\partial_z Y_A \
 =  \ \rho^3 \ T^{\rho \phi} \ , \qquad
&\partial_\rho Y_A \
 =  \ - \rho^3 \ T^{z \phi} \ , \nonumber \\
		&\Leftrightarrow  \qquad
T^{\rho \phi} \
 =  \ \frac{1}{\rho^3} \ \partial_z Y_A \ , \qquad
&T^{z \phi} \
 =  \ - \frac{1}{\rho^3} \ \partial_\rho Y_A \ .
	\label{TT FAC Y}
\end{eqnarray}
Then using \eref{TT FAC trT0} and substituting from \eref{TT FAC X} above, \eref{TT FAC DivTrho0} becomes:
\begin{eqnarray}
&0
 &= \partial_\rho T^{\rho \rho}
    + \partial_z T^{\rho z}
    + \frac{1}{\rho} \
      \left(T^{\rho \rho} + T^{z z}\right)
    + \frac{1}{\rho} \ T^{\rho \rho}
	\nonumber \\
 &&= \frac{1}{\rho^2} \ \partial_\rho
      \left(\rho^2 \ T^{\rho \rho}\right)
    + \frac{1}{\rho} \ \partial_{z z} X_A
    - \frac{1}{\rho^2} \ \partial_\rho X_A \ , \nonumber \\
	\nonumber \\
\Leftrightarrow \qquad
&&\partial_\rho \left(\rho^2 \ T^{\rho \rho}\right) \
 =  \partial_\rho X_A
    - \rho \ \partial_{z z} X_A \ , \nonumber \\
	\nonumber \\
\Leftrightarrow \qquad
&&T^{\rho \rho}
 =  \frac{1}{\rho^2} \int \left[
        \partial_\rho X_A - \rho \ \partial_{z z} X_A
      \right] d \rho
  + \frac{1}{\rho^2} \ f_A(z)
	\ , \label{TT FAC Trr, X}
\end{eqnarray}
again noting the function of integration, depending here on $z$, but not $\rho$ (or $\phi$).

Transverse, trace-free and axially-symmetric tensors in flat space can then be given, in cylindrical coordinates, by the matrix expression:
\begin{equation}\label{TT FAC}
\fl T^{a b} =
\left(
  \begin{array}{ccc}
    \frac{1}{\rho^2} \int \left[
       - \rho \ \partial_{z z} X_A
       + \partial_\rho X_A
       \right] d \rho
     & \frac{1}{\rho} \ \partial_z X_A
     & \frac{1}{\rho^3} \partial_z Y_A \\
  + \frac{1}{\rho^2} \ f_A(z)
   & & \\
   & & \\
    \frac{1}{\rho} \ \partial_z X_A
     & - \frac{1}{\rho} \ \partial_\rho X_A
     & - \frac{1}{\rho^3} \partial_\rho Y_A \\
   & & \\
    \frac{1}{\rho^3} \partial_z Y_A
     & - \frac{1}{\rho^3} \partial_\rho Y_A
     & \frac{1}{\rho^3} \ \partial_\rho X_A
     - \frac{1}{\rho^4} \ f_A(z) \\
    & & - \frac{1}{\rho^4} \int \left[
       - \rho \ \partial_{z z} X_A
       + \partial_\rho X_A
       \right] d \rho \\
  \end{array}
 \right) ,
\end{equation}
depending on the two scalar potentials $X_A$ and $Y_A$, and the function of integration $f_A(z)$.

\subsection{Spherical Coordinates}
\label{sec:TT FS}

The flat $3$-space line element, in spherical-polar coordinates $(r, \ \theta, \ \phi)$, is given by:
\begin{equation}
d l^2 \
  = \ d r^2
    + r^2 d \theta^2
    + r^2 \sin^2 \theta \ d \phi^2
	\ , \label{TT FS Metric}
\end{equation}
and the non-zero connection coefficients by:
\begin{equation}\label{TT FS Connections}
\hspace{-1.5cm} \eqalign{
\Gamma^r_{\theta \theta} \
 =& \ - r \ , \cr
\Gamma^r_{\phi \phi} \
 =& \ - r \ \sin^2 \theta \ ,}
\qquad
\eqalign{
\Gamma^\theta_{r \theta} \
 =& \ \Gamma^\theta_{\theta r} \
 =  \ \frac{1}{r} \ , \cr
\Gamma^\theta_{\phi \phi} \
 =& \ - \cos \theta \ \sin \theta \ ,}
\qquad
\eqalign{
\Gamma^\phi_{r \phi} \
 =  \ \Gamma^\phi_{\phi r} \
 =& \ \frac{1}{r} \ , \cr
\Gamma^\phi_{\theta \phi} \
 =  \ \Gamma^\phi_{\phi \theta} \
 =& \ \frac{\cos \theta}{\sin \theta} \ .}
\end{equation}
Imposing axial symmetry from \eref{TT FA partial phi Tab}, the conditions for the symmetric tensor $T^{a b}$ to be both transverse and trace-free are given by the equations:
\numparts
\begin{eqnarray}
\fl 0
  =
D_b T^{r b}
  = \partial_r T^{r r}
    + \partial_\theta T^{r \theta}
    + \cancel{\partial_\phi T^{r \phi}}
    + \frac{2}{r} \ T^{r r}
    - r \ T^{\theta \theta}
    - r \sin^2 \theta \ T^{\phi \phi}
    + \frac{\cos \theta}{\sin \theta} \ T^{r \theta}
	\label{TT FS DivTr0} \ , \\
\fl 0
  =
D_b T^{\theta b} \
  = \partial_r T^{\theta r}
    + \partial_\theta T^{\theta \theta}
    + \cancel{\partial_\phi T^{\theta \phi}}
    + \frac{\cos \theta}{\sin \theta} \ T^{\theta \theta}
    - \cos \theta \ \sin \theta \ T^{\phi \phi}
    + \frac{4}{r} \ T^{r \theta}
        \label{TT FS DivTth0} \ , \\
\fl 0
  =
D_b T^{\phi b} \
  = \partial_r T^{\phi r}
    + \partial_\theta T^{\phi \theta}
    + \cancel{\partial_\phi T^{\phi \phi}}
    + \frac{4}{r} \ T^{r \phi}
    + 3 \ \frac{\cos \theta}{\sin \theta} \ T^{\theta \phi}
        \label{TT FS DivTph0} \ , \\
\fl 0
  = T^{r r}
    + r^2 \ T^{\theta \theta}
    + r^2 \sin^2 \theta \ T^{\phi \phi}
      \label{TT FS trT0} \ .
\end{eqnarray}
\endnumparts

Beginning with \eref{TT FS DivTr0}, and removing the $T^{\theta \theta}$ and $T^{\phi \phi}$ terms with \eref{TT FS trT0}:
\begin{eqnarray}
&0 \
 &= \ \frac{3}{r} \ T^{r r}
    + \partial_r T^{r r}
    + \frac{\cos \theta}{\sin \theta} \ T^{r \theta}
    + \partial_\theta T^{r \theta}
	\nonumber \\
&&= \ \frac{1}{r^3} \ \partial_r
      \left(r^3 \ T^{r r}\right)
    + \frac{1}{\sin \theta} \ \partial_\theta
      \left(\sin \theta \ T^{r \theta}\right)
	\ , \nonumber \\
	\nonumber \\
\Leftrightarrow \qquad
&\partial_r
     &\left(r^3 \sin \theta \ T^{r r}\right) \
 =  \ \partial_\theta
      \left(- r^3 \sin \theta \ T^{r \theta}\right)
      \ , \label{TT FS Trr-Trt}
\end{eqnarray}
and by the equivalence of mixed partial derivatives, there must exist a scalar potential $V$ such that:
\begin{eqnarray}
&\partial_\theta V \
  = \ r^3 \sin \theta \ T^{r r} \ , \qquad
&\partial_r V \
  = \ - r^3 \sin \theta \ T^{r \theta} \ , \nonumber \\
\Leftrightarrow \qquad
&T^{r r} \
  = \ \frac{\partial_\theta V}{r^3 \sin \theta} \ , \qquad
&T^{r \theta} \
  = \ - \frac{\partial_r V}{r^3 \sin \theta} \ . \label{TT FS V,Trr Trt}
\end{eqnarray}

Now taking \eref{TT FS DivTph0}:
\begin{eqnarray}
&0 \
 &= \ \frac{4}{r} \ T^{r \phi}
    + \partial_r T^{r \phi}
    + 3 \frac{\cos \theta}{\sin \theta} \ T^{\theta \phi}
    + \partial_\theta T^{\theta \phi}
	\nonumber \\
&&= \ \frac{1}{r^4} \ \partial_r
      \left(r^4 \ T^{r \phi}\right)
    + \frac{1}{\sin^3 \theta} \ \partial_\theta
      \left(\sin^3 \theta \ T^{\theta \phi}\right) \ , \nonumber \\
	\nonumber \\
\Leftrightarrow \qquad
     &\partial_r
     &\left(- r^4 \sin^3 \theta \ T^{r \phi}\right) \
 =  \ \partial_\theta
      \left(r^4 \sin^3 \theta \ T^{\theta \phi}\right)
      \ , \label{TT FS Trp-Ttp}
\end{eqnarray}
and again, by the equivalence of mixed partial derivatives, there
must exist a scalar potential $W$ such that:
\begin{eqnarray}
&\partial_\theta W \
  = \ - r^4 \sin^3 \theta \ T^{r \phi} \ , \qquad
&\partial_r W \
  = \ r^4 \sin^3 \theta \ T^{\theta \phi} \ , \nonumber \\
\Leftrightarrow \qquad
&T^{r \phi} \
  = \ - \frac{\partial_\theta W}{r^4 \sin^3 \theta} \ , \qquad
&T^{\theta \phi} \
  = \ \frac{\partial_r W}{r^4 \sin^3 \theta} \ .
	\label{TT FS W,Trp Ttp}
\end{eqnarray}

Finally, using \eref{TT FS trT0} to simplify, \eref{TT FS DivTth0} gives:
\begin{eqnarray}
\fl &0 \
  = \ \partial_r T^{r \theta}
    + \partial_\theta T^{\theta \theta}
    + \frac{\cos \theta}{\sin \theta} \ T^{\theta \theta}
    + \frac{\cos \theta}{r^2 \sin \theta} \
      \left(T^{r r} + r^2 T^{\theta \theta}\right)
    + \frac{4}{r} \ T^{r \theta}
	\nonumber \\ \fl & \ \ \
  = \ 2 \frac{\cos \theta}{\sin \theta} \
      T^{\theta \theta}
    + \partial_\theta T^{\theta \theta}
    + \frac{4}{r} \ T^{r \theta}
    + \partial_r T^{r \theta}
    + \frac{\cos \theta}{r^2 \sin \theta} \ T^{r r} \ , \nonumber \\
	\fl \nonumber \\
\fl \Leftrightarrow \qquad
&\frac{1}{\sin^2 \theta} \ \partial_\theta
      \left(\sin^2 \theta \ T^{\theta \theta}\right) \
 = \ - \frac{1}{r^4} \ \partial_r
      \left(r^4 \ T^{r \theta}\right)
    - \frac{\cos \theta}{r^5 \sin^2 \theta} \
      \partial_\theta V \ , \nonumber \\
	\fl \nonumber \\
\fl \Leftrightarrow \qquad
&T^{\theta \theta} \
 =  \ \frac{1}{r^3 \sin^2 \theta} \ 
	\int \left[
        \sin \theta \ \partial_{r r} V
      + \frac{\sin \theta}{r} \ \partial_r V
      - \frac{\cos \theta}{r^2} \ \partial_\theta V
      \right] d \theta
    + \frac{g(r)}{r^3 \sin^2 \theta} \ ,
	\label{TT FS Ttt, V}
\end{eqnarray}
noting a function of integration, similar to the cylindrical coordinates, depending here on $r$ but not $\theta$ (or $\phi$).

Hence, transverse, trace-free and axially-symmetric tensors in flat space can be given, in spherical coordinates, by the matrix expression:
\begin{equation}
\fl T^{a b} =
\left(
  \begin{array}{ccc}
     \frac{1}{r^3 \sin \theta} \ \partial_\theta V
 & - \frac{1}{r^3 \sin \theta} \ \partial_r V
 & - \frac{1}{r^4 \sin^3 \theta} \ \partial_\theta W \\
& & \\
   - \frac{1}{r^3 \sin \theta} \ \partial_r V
 &   \frac{1}{r^3 \sin^2 \theta} \int \big[
       \sin \theta \ \partial_{r r} V
 &   \frac{1}{r^4 \sin^3 \theta} \ \partial_r W \\
 &   + \frac{\sin \theta}{r} \ \partial_r V
     - \frac{\cos \theta}{r^2} \  \partial_\theta V\big] d \theta & \\
 &   + \frac{1}{r^3 \sin^2 \theta} \ g(r) & \\
& & \\
   - \frac{1}{r^4 \sin^3 \theta} \ \partial_\theta W
 &   \frac{1}{r^4 \sin^3 \theta} \ \partial_r W
 & - \frac{1}{r^5 \sin^3 \theta} \ \partial_\theta V
   - \frac{1}{r^3 \sin^4 \theta} \ g(r) \\
 & & - \frac{1}{r^3 \sin^4 \theta} \int \big[
       \sin \theta \ \partial_{r r} V \\
 & & + \frac{\sin \theta}{r} \ \partial_r V
     - \frac{\cos \theta}{r^2} \ \partial_\theta V
     \big] d \theta \\
  \end{array}
 \right) , \label{TT FS}
\end{equation}
depending on the two scalar potentials $V$ and $W$, and the function of integration $g(r)$.

\subsection{Removing Integrals}

Since the tensor in cylindrical coordinates has a similar form to the linear symmetry case, a relation based on \eref{TT FL X,Y - R,S} has been found between the scalar function $X_A$ and a scalar potential $R_A$:
\begin{equation}
X_A \
 =  \ \partial_\rho R_A
    + \frac{1}{\rho} R_A
	\ , \label{TT FAC X(R)}
\end{equation}
giving a new form for an axially-symmetric ${T T}$ tensor in cylindrical coordinates:
\begin{equation}\label{TT FAC NoInt}
\fl T^{a b} =
\left(
  \begin{array}{ccc}
       - \frac{1}{\rho} \ \partial_{z z} R_A
       + \frac{1}{\rho^2} \ \partial_\rho R_A
     &   \frac{1}{\rho} \ \partial_{\rho z} R_A
       + \frac{1}{\rho^2} \ \partial_z R_A
     &   \frac{1}{\rho^3} \ \partial_z Y_A \\
       + \frac{1}{\rho^3} R_A
   & & \\
   & & \\
         \frac{1}{\rho} \ \partial_{\rho z} R_A
       + \frac{1}{\rho^2} \ \partial_z R_A
     & - \frac{1}{\rho} \ \partial_{\rho \rho} R_A
       - \frac{1}{\rho^2} \ \partial_\rho R_A
     & - \frac{1}{\rho^3} \ \partial_\rho Y_A \\
   & & \\
         \frac{1}{\rho^3} \ \partial_z Y_A
     & - \frac{1}{\rho^3} \ \partial_\rho Y_A
     &   \frac{1}{\rho^3} \ \partial_{\rho \rho} R_A
       + \frac{1}{\rho^3} \ \partial_{z z} R_A \\
   & & - \frac{1}{\rho^5} R_A \\
  \end{array}
 \right) \ ,
\end{equation}
with no integrals, and as a result, no function of integration. An expression for the potential $R_A$ in terms of $X_A$ can also be found from the general ordinary differential equation solution for \eref{TT FAC X(R)}:
\begin{equation}
R_A \
  = \ \frac{1}{\rho} \int \rho \ X_A \ d \rho
    + \frac{1}{\rho} \ h_A (z)
	\ , \label{TT FAC R(X)}
\end{equation}
noting the function of integration $h_A(z)$, similar to \eref{TT FL R(X)}.

To find an axially-symmetric TT tensor in \emph{spherical} coordinates, without the presence of an integral, a coordinate transformation can be carried out on \eref{TT FAC NoInt}. As with the linear case, the derivatives of the cylindrical potentials $R_A$ and $Y_A$ are found with respect to $r$ and $\theta$ using the chain rule:
\begin{eqnarray}
\partial_\rho R_A \
 =& \ \frac{\partial R_A}{\partial x^i}
      \frac{\partial x^i}{\partial \rho} \
 =  \ \sin \theta \ \partial_r R_A
    + \frac{1}{r} \ \cos \theta \ \partial_\theta R_A
	\ , \nonumber \\
\partial_z R_A \
 =& \ \frac{\partial R_A}{\partial x^i}
      \frac{\partial x^i}{\partial z} \
 = \ \cos \theta \ \partial_r R_A
   - \frac{1}{r} \ \sin \theta \ \partial_\theta R_A 
	\ , \nonumber \\
& \nonumber \\
\partial_{\rho \rho} R_A \
 =& \ \sin^2 \theta \ \partial_{r r} R_A
    + \frac{1}{r} \cos^2 \theta \ \partial_r R_A
    + 2 \frac{1}{r} \cos \theta \sin \theta \ \partial_{r \theta} R_A
	\nonumber \\ &
    - 2 \frac{1}{r^2} \cos \theta \sin \theta \ \partial_\theta R_A
    + \frac{1}{r^2} \cos^2 \theta \ \partial_{\theta \theta} R_A
	\ , \nonumber \\
& \nonumber \\
\partial_{\rho z} R_A \
 =& \ \cos \theta \sin \theta \ \partial_{r r} R_A
    - \frac{1}{r} \cos \theta \sin \theta \ \partial_r R_A
	\nonumber \\ &
    + \frac{1}{r} \cos^2 \theta \ \partial_{r \theta} R_A
    - \frac{1}{r} \sin^2 \theta \ \partial_{r \theta} R_A
	\nonumber \\ &
    - \frac{1}{r^2} \cos^2 \theta \ \partial_\theta R_A
    + \frac{1}{r^2} \sin^2 \theta \ \partial_\theta R_A
    - \frac{1}{r^2} \cos \theta \sin \theta \ \partial_{\theta \theta} R_A
	\ , \nonumber \\
& \nonumber \\
\partial_{z z} R_A \
 =& \ \cos^2 \theta \
      \partial_{r r} R_A
    + \frac{1}{r} \sin^2 \theta \
      \partial_r R_A
    - \frac{1}{r} 2 \cos \theta \ \sin \theta \
      \partial_{r \theta} R_A
	\nonumber \\ &
    + 2 \ \frac{1}{r^2} \cos \theta \ \sin \theta \
      \partial_\theta R_A
    + \frac{1}{r^2} \sin^2 \theta \
      \partial_{\theta \theta} R_A
	\ , \nonumber \\
& \nonumber \\
\partial_\rho Y_A \
 =& \ \frac{\partial Y_A}{\partial x^i}
      \frac{\partial x^i}{\partial \rho} \
 =  \ \sin \theta \ \partial_r Y_A
    + \frac{1}{r} \ \cos \theta \ \partial_\theta Y_A
	\ , \nonumber \\
\partial_z Y_A \
 =& \ \frac{\partial Y_A}{\partial x^i}
      \frac{\partial x^i}{\partial z} \
 =  \ \cos \theta \ \partial_r Y_A
    - \frac{1}{r} \ \sin \theta \ \partial_\theta Y_A
	\ .
\end{eqnarray}
The components of $T^{a b}$ are then transformed from cylindrical to spherical coordinates, giving the spherical components in terms of $R_A$ and $Y_A$:
\begin{eqnarray}
\fl 
T^{r r}
 &= \
    - \frac{1}{r^3 \sin \theta} \ \partial_{\theta \theta} R_A
    - \frac{\cos \theta}{r^3 \sin^2 \theta} \ \partial_\theta R_A
    + \frac{1}{r^3 \sin \theta} \ R_A
	\ ,
	\nonumber \\
\fl & \nonumber \\
\fl 
T^{\theta \theta}
 &= \ 
    - \frac{1}{r^3 \sin \theta} \ \partial_{r r} R_A
    - \frac{1}{r^4 \sin \theta} \ \partial_r R_A
    + \frac{\cos \theta}{r^5 \sin^2 \theta} \ \partial_\theta R_A
    + \frac{\cos^2 \theta}{r^5 \sin^3 \theta} \ R_A
        \ ,
	\nonumber \\
\fl & \nonumber \\
\fl 
T^{r \theta} \
 &= \ \frac{1}{r^3 \sin \theta} \ \partial_{r \theta} R_A
    + \frac{\cos \theta}{r^3 \sin^2 \theta} \ \partial_r R_A
    + \frac{\cos \theta}{r^4 \sin^2 \theta} \ R_A
        \ ,
        \nonumber \\
\fl & \nonumber \\
\fl 
T^{\phi \phi} \
 &= \ 
    + \frac{1}{r^3 \sin^3 \theta} \ \partial_{r r} R_A
    + \frac{1}{r^4 \sin^3 \theta} \ \partial_r R_A
    + \frac{1}{r^5 \sin^3 \theta} \ \partial_{\theta \theta} R_A
    - \frac{1}{r^5 \sin^5 \theta} \ R_A
        \ ,
        \nonumber \\
\fl & \nonumber \\
\fl 
T^{r \phi} \
 &= \ - \frac{\partial_\theta Y_A}{r^4 \sin^3 \theta}
        \ ,
	\qquad \qquad
T^{\theta \phi} \
  = \ \frac{\partial_r Y_A}{r^4 \sin^3 \theta}
        \ .
\end{eqnarray}
These then give an integral-free expression for an axially-symmetric TT tensor in spherical coordinates:
\begin{eqnarray}
\fl &T^{a b} \ = \nonumber \\
\fl &\left(
  \begin{array}{ccc}
    - \frac{1}{r^3 \sin \theta} \ \partial_{\theta \theta} R_A
  &   \frac{1}{r^3 \sin \theta} \ \partial_{r \theta} R_A
  & - \frac{1}{r^4 \sin^3 \theta} \ \partial_\theta Y_A
		\\
    - \frac{\cos \theta}{r^3 \sin^2 \theta} \ \partial_\theta R_A
  & + \frac{\cos \theta}{r^3 \sin^2 \theta} \ \partial_r R_A
    + \frac{\cos \theta}{r^4 \sin^2 \theta} \ R_A
  & 		\\
    + \frac{1}{r^3 \sin \theta} \ R_A
  & 
  & 		\\
 & & 		\\
      \frac{1}{r^3 \sin \theta} \ \partial_{r \theta} R_A
  & - \frac{1}{r^3 \sin \theta} \ \partial_{r r} R_A
    - \frac{1}{r^4 \sin \theta} \ \partial_r R_A
  &   \frac{1}{r^4 \sin^3 \theta} \ \partial_r Y_A
		\\
    + \frac{\cos \theta}{r^3 \sin^2 \theta} \ \partial_r R_A
  & + \frac{\cos \theta}{r^5 \sin^2 \theta} \ \partial_\theta R_A
    + \frac{\cos^2 \theta}{r^5 \sin^3 \theta} \ R_A
  &  		\\
    + \frac{\cos \theta}{r^4 \sin^2 \theta} \ R_A
  &
  & 		\\
 & & 		\\
    - \frac{1}{r^4 \sin^3 \theta} \ \partial_\theta Y_A
  &   \frac{1}{r^4 \sin^3 \theta} \ \partial_r Y_A
  & \hspace{-1cm}
    + \frac{1}{r^3 \sin^3 \theta} \ \partial_{r r} R_A
    + \frac{1}{r^4 \sin^3 \theta} \ \partial_r R_A
		\\
  &
  & \hspace{-1cm}
    + \frac{1}{r^5 \sin^3 \theta} \ \partial_{\theta \theta} R_A
    - \frac{1}{r^5 \sin^5 \theta} \ R_A
		\\
  \end{array}
 \right)
	. \label{TT FS NoInt}
\end{eqnarray}
depending on $R_A$ and $Y_A$ alone, with no additional functions. This can be compared with \eref{TT FS} to show an equivalence between the cylindrical potential $Y_A$ and the spherical potential $W$. It proves a little more difficult however, to find a direct relation between $V$ and either $R_A$ or $X_A$.

\section{General Space TT Tensors}
\label{sec:TTT General Space}

In this section, the flat space metric is replaced by a more general spatial metric with an axial symmetry. The conditions for an axially-symmetric two-tensor to be transverse and trace-free are given, but only one scalar potential can be found using the techniques of the previous two sections, coinciding with  that given in \cite{BakerPuzio} and \cite{Dain}. The remaining equations are reduced to a second order partial differential equation in two of the components.

\subsection{Metric and Tensors with Axial Symmetry}
\label{sec:TT Brill}

A general axially symmetric $3$-space metric can be given by the Brill wave metric (3) of \cite{Brill}, which was credited there to H. Bondi. This is geometrically equivalent to the metric used in \cite{BakerPuzio}, with a related set of functions. In cylindrical-polar type coordinates $(\rho, \ z, \ \phi)$, the conformal part of this metric can be expressed as:
\begin{equation}\label{TT GA Metric}
d l^2 \
  = \ e^{2 q} d r^2
    + e^{2 q} d z^2
    + \rho^2 d \phi^2 \ ,
\end{equation}
for a differential function $q(\rho, z)$ such that:
\begin{equation}\label{TT GA q Def}
     q \ |_{\rho = 0} \
 = \ \partial_\rho \ q \ |_{\rho = 0} \
 = \ 0 \ ,
\end{equation}
and that $q$ decays faster than $\frac{1}{r}$ at infinity, and is reasonably differential. The non-zero connection coefficients for this metric are then given by:
\begin{equation}\label{TT GA Connections}
\eqalign{
\Gamma^\rho_{\rho \rho}
 =  \partial_\rho \ q \ , \cr
\Gamma^\rho_{\rho z}
 =  \Gamma^\rho_{z \rho}
 =  \partial_z \ q \ , \cr
\Gamma^\rho_{z z}
 =  - \partial_\rho \ q \ , \cr
\Gamma^\rho_{\phi \phi}
 =  - \rho \ e^{- 2 q} \ ,}
 \qquad
\eqalign{
\Gamma^z_{\rho \rho}
 =  - \partial_z \ q \ , \cr
\Gamma^z_{\rho z}
 =  \Gamma^z_{z \rho}
 =  \partial_\rho \ q \ , \cr
\Gamma^z_{z z}
 =  \partial_z \ q \ , \cr}
 \qquad
\eqalign{
 \cr
\Gamma^\phi_{\rho \phi}
 =  \Gamma^\phi_{\phi \rho}
 =  \frac{1}{\rho} \ , \cr
 \cr}
\end{equation}
and the conditions for a symmetric tensor $T^{a b}$ to be transverse and trace-free with respect to \eref{TT GA Metric}, along with the condition of axial-symmetry \eref{TT FA partial phi Tab}, are given by the equations:
\numparts
\begin{eqnarray}
\fl 0
 =    D_b T^{\rho b}
 =   \partial_\rho T^{\rho \rho}
    + \partial_z T^{\rho z}
    + \cancel{\partial_\phi T^{\rho \phi}}
    + (3 \ \partial_\rho q
      + \frac{1}{\rho}) \ T^{\rho \rho}
    + 4 \ \partial_z q \ T^{\rho z}
        \nonumber \\
    - \partial_\rho q \ T^{z z}
    - \rho \ e^{- 2 q} \ T^{\phi \phi}
        \label{TT GA divTr} \ ,
        \\
\fl 0
 =    D_b T^{z b}
 =    \partial_\rho T^{z \rho}
    + \partial_z T^{z z}
    + \cancel{\partial_\phi T^{z \phi}}
    - \partial_z q \ T^{\rho \rho}
    + 3 \ \partial_z q \ T^{z z}
    + (4 \ \partial_\rho q
      + \frac{1}{\rho}) \ T^{\rho z}
        \label{TT GA divTz} \ ,
        \\
\fl 0
 =    D_b T^{\phi b}
 =    \partial_\rho T^{\phi \rho}
    + \partial_z T^{\phi z}
    + \cancel{\partial_\phi T^{\phi \phi}}
    + (2 \ \partial_\rho q
      + \frac{3}{\rho}) \ T^{\rho \phi}
    + 2 \ \partial_z q \ T^{z \phi}
        \label{TT GA divTp} \ ,
        \\
\fl 0 
 =   e^{2 q} \ T^{\rho \rho}
    + e^{2 q} \ T^{z z}
    + \rho^2 \ T^{\phi \phi}
        \ .
        \label{TT GA trT}
\end{eqnarray}
\endnumparts

Working from \eref{TT GA divTp}:
\begin{eqnarray}
\fl &0 \
 &= \ e^{2 q} \ \partial_\rho T^{\rho \phi}
    + T^{\rho \phi} \ \partial_\rho e^{2 q}
    + T^{\rho \phi} \ e^{2 q} \ \frac{1}{\rho^3} \partial_\rho \rho^3
    + \ e^{2 q} \ \partial_z T^{z \phi}
    + T^{z \phi} \ \partial_z e^{2 q}
        \nonumber \\ \fl &
 &= \ \frac{1}{\rho^3} \
      \partial_\rho \ (\rho^3 \ e^{2 q} \ T^{\rho \phi})
    + \frac{1}{\rho^3} \
      \partial_z \ (\rho^3 \ e^{2 q} \ T^{z \phi})
        \ , \nonumber \\
\fl \nonumber \\
\fl \Leftrightarrow \qquad
&\partial_\rho \ &(\rho^3 \ e^{2 q} \ T^{\rho \phi}) \
 =  \ \partial_z \ (- \rho^3 \ e^{2 q} \ T^{z \phi})
        \ ,
\end{eqnarray}
and the equivalence of mixed partial derivatives implies the existence of a scalar potential $w$ such that:
\begin{eqnarray}
&\partial_z \ w \
  = \ \rho^3 \ e^{2 q} \ T^{\rho \phi}
	\ , \qquad
&\partial_\rho \ w \
  = \ - \rho^3 \ e^{2 q} \ T^{z \phi}
	\ , \nonumber \\
\nonumber \\
\Leftrightarrow \qquad
&T^{\rho \phi}
  = \rho^{- 3} e^{- 2 q} \partial_z w
	\ , \qquad
&T^{z \phi}
  = - \rho^{- 3} e^{- 2 q} \partial_\rho w
	\ . \label{TT GA Potential}
\end{eqnarray}
With the space-time assumed to have a ``time-rotation'' symmetry, these components are shown in \cite{BrandtSeidel} to give the only non-zero components of the extrinsic curvature. In this case, the tensor agrees exactly with the curvature given by \cite{Dain}. The components given by \eref{TT GA Potential} can also be seen to agree with the corresponding curvature components of \cite{BakerPuzio}, and if $q = 0$, i.e. the metric is reduced to a flat space metric, \eref{TT GA Potential} is also equivalent to \eref{TT FAC Y} from \sref{sec:TT FAC}.

\subsection{Equations for Remaining Tensor Components}
\label{sec:TT GA Other Eqns}

Manipulating both \eref{TT GA divTr} and \eref{TT GA divTz} similar to section \eref{sec:TT FAC}, using \eref{TT GA trT} to remove the $T^{\phi \phi}$ component from \eref{TT GA divTr}:
\begin{eqnarray}
0 \
 =& \ \frac{1}{\rho^2} \ \partial_\rho \
      (\rho^2 \ e^{2 q} \ T^{\rho \rho})
    + \frac{1}{\rho^2} \ \partial_z \
      (\rho^2 \ e^{2 q} \ T^{\rho z})
    + \frac{1}{2} \ T^{\rho \rho} \
      \partial_\rho e^{2 q}
	\nonumber \\ &
    + T^{\rho z} \ \partial_z e^{2 q}
    - \frac{1}{2} \ \rho^2 \ T^{z z} \
      \partial_\rho \ (\rho^{- 2} e^{2 q})
	\ , \label{TT GA Eq Trr-Tzr} \\
	\nonumber \\
0 \
 =& \ \frac{1}{\rho^2} \ \partial_\rho \
      (\rho^2 \ e^{2 q} \ T^{\rho z})
    + \frac{1}{\rho^2} \ \partial_z \
      (\rho^2 \ e^{2 q} \ T^{z z})
    + T^{\rho z} \partial_\rho e^{2 q}
	\nonumber \\ &
    - \frac{1}{2 \rho^2} \ e^{2 q} \ T^{\rho z} \
      \partial_\rho \rho^2
    + \frac{1}{2} \ T^{z z} \ \partial_z e^{2 q}
    - \frac{1}{2} \ T^{\rho \rho} \
      \partial_z e^{2 q}
        \ , \label{TT GA Eq Tzz-Tzr}
\end{eqnarray}
where there remains three tensor components in each equation, rather than the two required for the techniques used previously. However both equations do contain similar terms for the $T^{\rho z}$ component. Taking \eref{TT GA Eq Tzz-Tzr} first, and bringing all of the $T^{\rho z}$ terms to one side:
\begin{eqnarray}
      \partial_\rho& \
      (\rho^2 \ e^{2 q} \ T^{\rho z})
    + \rho^2 \ T^{\rho z} \ \partial_\rho e^{2 q}
    - \frac{1}{2} \ e^{2 q} \ T^{\rho z} \
      \partial_\rho \rho^2
        \nonumber \\
 =& \ - \partial_z \
      (\rho^2 \ e^{2 q} \ T^{z z})
    - \frac{1}{2} \ \rho^2 \ T^{z z} \ \partial_z e^{2 q}
    + \frac{1}{2} \ \rho^2 \ T^{\rho \rho} \
      \partial_z e^{2 q}
        \ .
        \label{TT GA Eq Tzz-Tzr 2}
\end{eqnarray}
Since the $T^{\rho z}$ terms have derivatives with respect to $\rho$ here, and $z$ in \eref{TT GA Eq Trr-Tzr}, both sides are integrated with respect to $\rho$:
\begin{eqnarray}
      \rho^2 \, e^{2 q} \, T^{\rho z}
    + \int \left(\rho^2 \, T^{\rho z}\right) d e^{2 q}
    - \frac{1}{2} \int \left(
        e^{2 q} \, T^{\rho z}\right) d \rho^2 \ =
        \nonumber \\
\fl - \int \partial_z
      (\rho^2 \, e^{2 q} \, T^{z z}) d \rho
    - \frac{1}{2} \int \left(
        \rho^2 \, T^{z z} \, \partial_z e^{2 q}
      \right) d \rho
    + \frac{1}{2} \int \left(
        \rho^2 \, T^{\rho \rho} \,
        \partial_z e^{2 q}
      \right) d \rho
    + f_I (z)
        \, .
        \label{TT GA Eq Tzz-Tzr 3}
\end{eqnarray}

Taking now \eref{TT GA Eq Trr-Tzr}, and again bringing the $T^{\rho z}$ terms to one side:
\begin{eqnarray}
      \frac{1}{\rho^2} \ \partial_z \
      (\rho^2 e^{2 q} \ T^{\rho z})
    + T^{\rho z} \ \partial_z e^{2 q}
        \nonumber \\
 = \ - \frac{1}{\rho^2} \ \partial_\rho \
      (\rho^2 e^{2 q} \ T^{\rho \rho})
    - \frac{1}{2} \ T^{\rho \rho} \
      \partial_\rho e^{2 q}
    + \frac{1}{2} \ \rho^2 \ T^{z z} \ \partial_\rho \
      (\rho^{- 2} e^{2 q})
        \ .
        \label{TT GA Eq Trr-Tzr 2}
\end{eqnarray}
With both sides of \eref{TT GA Eq Trr-Tzr 2} integrated with respect to $z$, there is still a difference with \eref{TT GA Eq Tzz-Tzr 3}, however the missing term can be given by first adding a term involving the derivative $\partial_z \rho^2$, which itself evaluates to zero. Hence \eref{TT GA Eq Trr-Tzr 2} is equivalent to:
\begin{eqnarray}
      \rho^2 e^{2 q} \, T^{\rho z}
    + \int \left(
      \rho^2 \, T^{\rho z} \right) d e^{2 q}
    - \frac{1}{2} \int \left(
      e^{2 q} \, T^{\rho z} \right) d \rho^2
        \ =
        \nonumber \\
\fl - \int \partial_\rho
      (\rho^2 \, e^{2 q} \, T^{\rho \rho}) d z
    - \frac{1}{2} \int \left(
        \rho^2 \, T^{\rho \rho} \, \partial_\rho e^{2 q}
      \right) d z
    + \frac{1}{2} \int \left(
        \rho^4 \, T^{z z} \,
        \partial_\rho (\rho^{- 2} e^{2 q})
      \right) d z
    + f_J (\rho)
        \, .
        \label{TT GA Eq Trr-Tzr 3}
\end{eqnarray}

Since the left hand sides of both \eref{TT GA Eq Tzz-Tzr 3} and \eref{TT GA Eq Trr-Tzr 3} are equivalent, the right hand sides can be equated to give a single equation, depending on $T^{\rho \rho}$ and $T^{z z}$ alone:
\begin{eqnarray}
\fl - \int \partial_z
      (\rho^2 \, e^{2 q} \, T^{z z}) d \rho
    - \frac{1}{2} \int \left(
      \rho^2 \, T^{z z} \, \partial_z e^{2 q}\right) d \rho
    + \frac{1}{2} \int \left(
      \rho^2 \, T^{\rho \rho} \, \partial_z e^{2 q}\right)
      d \rho
    + f_I (z) \ =
        \nonumber \\
\fl - \int \partial_\rho
      (\rho^2 \, e^{2 q} \, T^{\rho \rho}) d z
    - \frac{1}{2} \int \left(
      \rho^2 \, T^{\rho \rho} \, \partial_\rho e^{2 q}
      \right) d z
    + \frac{1}{2} \int \left(
      \rho^4 \, T^{z z} \, \partial_\rho (\rho^{- 2} e^{2 q})
      \right) d z
    + f_J (\rho)
        \, .
\end{eqnarray}
Differentiating both sides with respect to both $\rho$ and $z$, gives a second order partial differential equation in $T^{\rho \rho}$ and $T^{z z}$:
\begin{eqnarray}
  \partial_z \partial_z \
      (\rho^2 e^{2 q} \ T^{z z})
    + \frac{1}{2} \partial_z \left(
        \rho^2 \ T^{z z} \ \partial_z \ e^{2 q}
      \right)
    - \frac{1}{2} \partial_z \left(
        \rho^2 \ T^{\rho \rho} \ \partial_z \ e^{2 q}
      \right) \ =
        \nonumber \\
      \partial_\rho \partial_\rho \
      (\rho^2 e^{2 q} \ T^{\rho \rho})
    + \frac{1}{2} \partial_\rho \left(
        \rho^2 \ T^{\rho \rho} \
        \partial_\rho \ e^{2 q}
      \right)
    - \frac{1}{2} \partial_\rho \left(
        \rho^4 \ T^{z z} \
        \partial_\rho \ (\rho^{- 2} e^{2 q})
      \right) .
	\label{TT GA Eq Final 1}
\end{eqnarray}
Unfortunately, finding a relation between the components $T^{\rho \rho}$ and $T^{z z}$ involves solving this equation, requiring two separate boundary conditions to give a unique solution. Assuming enough information is available to solve \eref{TT GA Eq Final 1}, the relation between $T^{\rho \rho}$ and $T^{z z}$ can be used, along with any boundary conditions, to solve either \eref{TT GA Eq Trr-Tzr} or \eref{TT GA Eq Tzz-Tzr} for a relation involving $T^{\rho z}$. The full tensor can then be found using \eref{TT GA trT}, depending on the potential $w$, and a potential obtained from the relations between $T^{\rho \rho}$, $T^{z z}$ and $T^{\rho z}$.

\section{Potentials for Bowen-York Curvature}
\label{sec:TTT BY}

Since the Bowen-York conformal extrinsic curvature \cite{BY} is transverse, trace-free and axially-symmetric in a conformally flat space, it should be given by an appropriate choice of potentials for the tensors derived in \sref{sec:TTT Flat Space Axial}.

\subsection{Angular Momentum Part}

Taking first the case of zero linear momentum, the conformal Bowen-York curvature is given by:
\begin{equation}\label{TT BY Ang Cyl}
\cK_{a b} \
 =  \ \frac{3}{r^3} \ \left(
        \epsilon_{a c d} \ q_b
      + \epsilon_{b c d} \ q_a\right)
      q^c J^d
        \ ,
\end{equation}
depending on the angular momentum $J^a$ alone, with $q^a$ the unit normal to a sphere of constant radius, and $\epsilon_{a b c}$ the Levi-Civita alternating tensor. In cylindrical coordinates $(\rho, \ z, \ \phi)$, with the angular momentum directed in the axial direction, i.e., along the positive $z$ coorindate, the vectors $q^a$ and $J^a$ are given by:
\begin{equation}
q^a \
 =  \ q_a \
 =  \ \frac{(\rho, \ z, \ 0)}{\sqrt{\rho^2 + z^2}} \ ,
  \qquad
J^a \
 =  \ (0, \ J, \ 0) \ .
\end{equation}
Recalling from \sref{sec:TT FAC}, that the Levi-Civita tensor has a reversed sign for the coordinates given in this order, the non-zero terms of the conformal Bowen-York curvature are given, in cylindrical coordinates, by:
\begin{eqnarray}\label{TT BY Ang Cyl Lower}
\cK_{\rho \phi} \
 =  \ \cK_{\phi \rho} \
 =  \ \frac{3 J \rho^3}{r^5} \ , \qquad
\cK_{z \phi} \
 =  \ \cK_{\phi z} \
 =  \ \frac{3 J \rho^2 z}{r^5} \ .
\end{eqnarray}

To find the necessary choice of potentials, the components of the tensor expressions \eref{TT FAC} and \eref{TT FAC NoInt} are set equal to those of the Bowen-York curvature \eref{TT BY Ang Cyl} with indices raised, both giving:
\begin{eqnarray}
&\partial_z Y_A \
 =  \ \frac{3 J \rho^4}{r^5}
          \ ,
        \qquad
\partial_\rho Y_A \
 =  \ - \frac{3 J \rho^3 z}{r^5}
          \ ,
          \label{TT BY Ang v T cyl}
\end{eqnarray}
with the remaining tensor components equal to zero, implying the potentials $X_A$, $R_A$ must be constants. Integrating the two equations in \eref{TT BY Ang v T cyl}, with respect to $z$ and $\rho$ respectively, leads to the solution:
\begin{eqnarray}\label{TT BY Ang Integral Soln}
Y_A \ = \ J \ \frac{3 \rho^2 z + 2 z^3}{r^3} \ ,
\end{eqnarray}
plus a constant of integration, which is differentiated out when $Y_A$ is used to produce a TT tensor. Since $Y_A$ is equivalent to the spherical potential $W$, and the remaining potential must be a constant, \eref{TT BY Ang Integral Soln} can easily be translated into spherical coordinates, giving:
\begin{eqnarray}
W = - J (3 \sin^2 \theta \ \cos \theta + 2 \cos^3 \theta)
  = J (\cos^3 \theta - 3 \cos \theta)
          \, ,
          \label{TT BY Ang Sph}
\end{eqnarray}
which agrees with (21) of \cite{DainLT}, for the curvature tensor derived in \cite{Dain}.

\subsection{Linear Momentum Part}

Taking now, the \emph{angular} momentum to be zero, the conformal Bowen-York extrinsic curvature is given by:
\begin{eqnarray}\label{TT BY Lin Cyl}
\cK_{a b}^\pm \
 &= \ \frac{3}{2 r^2} \ \left[
        P_a q_b + P_b q_a
      - (\cg_{a b} - q_a q_b) P^c q_c \right]
        \nonumber \\ &
  \mp \frac{3 a^2}{2 \ r^4} \left[
        P_a q_b + P_b q_a
     + (\cg_{a b} - 5 q_a q_b) P^c q_c \right]
        \ ,
\end{eqnarray}
where $P^a$ denotes the linear momentum of a single source, $q^a$ the unit normal to a sphere of constant radius and $a$ an arbitrary constant. This can only give an axially-symmetric tensor, if the momentum is directed along the axis. Hence, in cylindrical type coordinates $(\rho, \ z, \ \phi)$, the linear momentum vector and unit space-like normal $q^a$ are given by:
\begin{equation}
P^a \
 =  \ (0, \ P, \ 0)
      \ ,
  \qquad
q^a \
 =  \ q_a \
 =  \ \frac{(\rho, \ z, \ 0)}{\sqrt{\rho^2 + z^2}}
      \ ,
\end{equation}
giving the non-zero components of the conformal Bowen-York curvature as:
\numparts
\begin{eqnarray}\label{TT BY Lin Cyl Lower}
\cK_{\rho \rho}^\pm \
 =  \ \frac{3 P z}{2 \ r^5} (- r^2 + \rho^2) \
  \mp \ \frac{3 a^2 P z}{2 \ r^7} (r^2 - 5 \ \rho^2)
          \ , \\
\cK_{\rho z}^\pm \
 =  \ \frac{3 P \rho}{2 \ r^5} (r^2 + z^2) \
  \mp \ \frac{3 a^2 P \rho}{2 \ r^7} (r^2 - 5 \ z^2)
          \ , \\
\cK_{z z}^\pm \
 =  \ \frac{3 P z}{2 \ r^5} (r^2 + z^2) \
  \mp \ \frac{3 a^2 P z}{2 \ r^7} (3 \ r^2 - 5 \ z^2)
          \ , \\
\cK_{\phi \phi}^\pm \
 =  \ - \frac{3 P \rho^2 z}{2 \ r^3} \
  \mp \ \frac{3 a^2 P \rho^2 z}{2 \ r^5}
          \ ,
\end{eqnarray}
\endnumparts
with $r = \sqrt{\rho^2 + z^2}$, and the vanishing components showing the potential $Y_A$ to be a constant.

Working firstly with the potential $X_A$ and the expression \eref{TT FAC}, the ``$\rho z$'' components are equated with those of \eref{TT BY Lin Cyl Lower} with indices raised:
\begin{eqnarray}
X_A \
 =  \ \frac{3}{2} P \rho^2 \int \frac{1}{r^5} \
      (2 \rho^2 + z^2) d z
  \mp \ \frac{3}{2} a^2 P \rho^2 \int \frac{1}{r^7} \
      (\rho^2 - 4 z^2) d z
          \ ,
          \label{TT BY Lin v T cyl 1}
\end{eqnarray}
and equating next the ``$z z$'' components:
\begin{eqnarray}
X_A \
 = \  - \frac{3}{2} P z \int \frac{\rho}{r^5} \
       (2 \rho^2 + z^2) d \rho
  \pm \frac{3}{2} a^2 P z \int \frac{\rho}{r^7} \
       (3 \rho^2 - 2 z^2) d \rho
          \ .
          \label{TT BY Lin v T cyl 2}
\end{eqnarray}
Carrying out the two sets of integrals, noting that $r = \sqrt{\rho^2 + z^2}$, gives:
\begin{equation}
X_A \ = \ P \ \frac{3 \rho^2 z + 4 z^3}{2 r^3} \
  \mp \ P a^2 \frac{3 \ \rho^2 z}{2 r^5}
	\ . \label{BY X}
\end{equation}
A constant of integration can also be added, but as with \eref{TT BY Ang Integral Soln}, it does not have an influence on the tensor. To find the corresponding expression for the potential $R_A$ of \eref{TT FAC NoInt}, the relation \eref{TT FAC R(X)} between $R_A$ and $X_A$ is used:
\begin{eqnarray}
R_A \
 &= \ \frac{1}{\rho} \int \rho \ X_A \ d \rho
	\nonumber \\
 &= \ \frac{1}{\rho} \int 
        P \ \frac{3 \rho^3 z + 4 \rho z^3}{2 r^3}
      \ d \rho \
\mp \ \frac{1}{\rho} \int
        P a^2 \frac{3 \ \rho^3 z}{2 r^5}
      \ d \rho \ .
\end{eqnarray}
Although functions of integration of the form:
\begin{equation}
\frac{1}{\rho} \int
        \rho \ k \ d \rho
    + \frac{1}{\rho} \ h_A (z) \ ,
\end{equation}
can also be added, where $k$ is the constant of integration from \eref{BY X}, these will all be cancelled or differentiated out when forming a tensor using \eref{TT FAC NoInt}.

The two solutions for the Bowen-York curvature can be combined, with both momenta directed along the $z$-axis, and the choice of scalar potentials for \eref{TT FAC} and \eref{TT FAC NoInt} given by:
\begin{eqnarray}
\fl &X_A \
 =  \ P \ \frac{3 \rho^2 z + 4 z^3}{2 r^3} \
  \mp \ P a^2 \frac{3 \ \rho^2 z}{2 r^5} \ , \qquad
&Y_A \
 =  \ J \ \frac{3 \rho^2 z + 2 z^3}{r^3} \ .
\nonumber \\
\fl &R_A \
 =  \ \frac{1}{\rho} \int 
        P \ \frac{3 \rho^3 z + 4 \rho z^3}{2 r^3}
      \ d \rho \
\mp \ \frac{1}{\rho} \int
        P a^2 \frac{3 \ \rho^3 z}{2 r^5}
      \ d \rho \ , \qquad&
\label{TT BY Combined Potentials}
\end{eqnarray}
Due to the relation with the Bowen-York curvature, in general $X_A$, $R_A$ can be considered to represent a linear momentum and $Y_A$ an angular momentum, when used with \eref{TT FAC}, \eref{TT FAC NoInt} for an axially-symmetric TT tensor.

\section{Conclusion}

In flat $3$-space, expressions have successfully been given for transverse trace-free tensors in different coordinate systems, with both linear and axial symmetries, depending on only two scalar potentials. In a more general axially-symmetric $3$-space, a single potential has been derived, equivalent to \cite{BakerPuzio} and \cite{Dain}, for two of the components. For the remaining components, the TT conditions have been reduced to a second order partial differential equation, requiring boundary conditions to be solved. The axially-symmetric flat space tensors have also been compared with the Bowen-York curvature, and specific choices of the potentials shown to give the Bowen-York curvature. There is also a distinct relationship between each of the potentials, and either the angular or linear momentum of the Bowen-York space. The expressions derived could however benefit from a coordinate independent form, in line with that given in \cite{Dain}. An expanded version of some of the content can be found in \cite{RoryPhD}.

\ack

This research was supported by SFI grant no. 07/RFP/PHYF148. We would also like to thank Prof. Robert Beig and Dr. Xie Naqing, for some very helpful comments.

\section*{References}

\bibliography{TTTensors}
\bibliographystyle{unsrt}

\end{document}